\documentclass{article}
\usepackage{graphicx}

\title{Improving Retrieval for RAG based Question Answering Models on Financial Documents}
\author{Spurthi Setty, Harsh Thakkar, Alyssa Lee, Eden Chung, Natan Vidra }
\date{}

\begin{document}

\maketitle

\begin{abstract} 
\noindent 
The effectiveness of Large Language Models (LLMs) in generating accurate responses relies heavily on the quality of input provided, particularly when employing Retrieval Augmented Generation (RAG) techniques. RAG enhances LLMs by sourcing the most relevant text chunk(s) to base queries upon. Despite the significant advancements in LLMs' response quality in recent years, users may still encounter inaccuracies or irrelevant answers; these issues often stem from suboptimal text chunk retrieval by RAG rather than the inherent capabilities of LLMs. To augment the efficacy of LLMs, it is crucial to refine the RAG process. This paper explores the existing constraints of RAG pipelines and introduces methodologies for enhancing text retrieval. It delves into strategies such as sophisticated chunking techniques, query expansion, the incorporation of metadata annotations, the application of re-ranking algorithms, and the fine-tuning of embedding algorithms. Implementing these approaches can substantially improve the retrieval quality, thereby elevating the overall performance and reliability of LLMs in processing and responding to queries.

\end{abstract}

\section{Introduction}
    In recent years, the emergence of Large Language Models (LLMs) represent a critical turning point in Generative AI and its ability to expedite productivity across a variety domains. However, the capabilities of these models, while impressive, are limited in a number of ways that have hindered certain industries from being able to take full advantage of the potential of this technology. A key disadvantage is the tendency for LLMs to hallucinate information and its lack of knowledge in domain specific areas. The knowledge of LLMs are limited by their training data, and without the use of additional techniques, these models have very poor performance of very domain specific tasks. 
    \\
    
    \noindent In order to develop a large language model, the first step is the pre-training process where a transformer is trained on a very large corpus of text data. This data is very general and not specific to a certain domain or field, as well as unchanging with time. This is a reason why LLMs like ChatGPT might perform well for general queries but fail on questions on more specific and higher-level topics. Additionally, a model's performance about a certain topic is highly dependent on how often that information appears in the training data, meaning that LLMs struggle with information that does not appear frequently. \footnote{Shahul Es, Jithin James, Luis Espinosa-Anke, Steven Schockaert. RAGAS: Automated Evaluation of Retrieval Augmented Generation, 2023. arXiv:2309.15217} This is the case for most domain-specific information, such as financial information and expertise, which is why standard LLMs have poor performance with domain-specific questions.
    \\
    
    \noindent The concept of knowledge injection refers to the ability to give the model access to information beyond its original training data, and the main way we can improve the performance of large language models on domain-specific tasks. The two primary techniques of knowledge injection are additional training of the model or fine-tuning, or in-context learning, the most popular version of which is Retrieval Augmented Generation (RAG). 
    
    \subsection{Fine-tuning}

    Fine-tuning is the process of updating model parameters for specific tasks and domains. Through fine-tuning one can further train a LLM on a domain-specific dataset so that it has access to additional information not present in the original training data. Fine-tuning also enables us to train the model to get consistent outputs via instruction fine-tuning. The three main methods of fine-tuning are unsupervised fine-tuning, supervised fine-tuning, and reinforcement learning. In our previous work, we examined the method of using supervised fine-tuning to improve several different models on finance-specific questions. Supervised fine-tuning, particularly instruction fine-tuning is ideal to tailor a model to perform a certain way given that it has access to labelled data with input and output pairs. However, a primary bottleneck of this method is access to high-quality data in order to conduct effective supervised fine-tuning. For question-answering tasks, these datasets are usually curated with human experts, often with the aid of other LLMs to get accurate and comprehensive question-answer pairs. 
    \\
    
    \noindent In unsupervised fine-tuning, the data is unlabeled and can be viewed as an extension of the pre-training process, but with domain-specific, instead of general documents and text. However, the model is not given any additional instructions or labels from this data. The hope is that the model will retain the information from this additional training data and be able to access that knowledge when doing very domain-specific question-answering. While this approach of fine-tuning can be much more scalable compared to supervised fine-tuning because of the availability of data, the results of the fine-tuned model won't be as robust. Moreover, while domain-specific data is used, there is no guarantee that the LLM will be able to access it when it is needed, especially if the size of the domain-specific data is small when compared to the text-corpus that was used for pre-training.

\subsection{Retrieval Augmented Generation}   
    Retrieval Augmented Generation is a method of in-context learning that allows Large Language Models to have access to new knowledge sources and answer questions. In context learning is a method of knowledge injection where one improves pre-trained LLMs by chaining the input query with additional knowledge. \footnote{Oded Ovadia, Menachem Brief, Moshik Mishaeli, and Oren Elisha.
    Fine-Tuning or Retrieval? Comparing Knowledge Injection in LLMs, 2023. arXiv:2312.05934.} It does not involve updating any of the model weights like fine-tuning, but rather changing the query to include additional information the model can refer to it. A benefit of this is that because the knowledge is being fed directly, the LLM can refer to it directly and there is no concern of this knowledge getting lost in the training data like it might be via unsupervised fine-tuning. However, due to the limited context window of large language models, there is only so much information that can be chained to a query, and it is often much shorter than the total amount of knowledge we want to augment the model with. 
    \\
    
    \noindent Thus comes in the retrieval aspect of retrieval augmented generation (RAG), in order to decide which chunk of text to add to the input query for question-answering tasks. We explore the use case of document-based question answering, specifically for finance-specific documents and queries. Suppose one has several reports such as 10-ks or earning transcripts and would like to chat or ask questions on them. For this use case, we need to somehow effectively inject the knowledge of those financial documents into the large language model. In order to prevent hallucinations, some sort of in-context learning approach would be ideal where we can specify via prompts for the model to only refer to the additional context given with the question. For example by the following as a systems prompt: 
    """You are a financial chatbot trained to answer questions based on the information provided in 10-K
    documents. Your responses should be directly sourced from the content of these documents."""
    and this as a user prompt: 
    f"Context: {context} Query: {query} Answer:"
    \\
    
    \noindent Due to the size and complexity of the documents we want to chat with, it is not possible to simply add the entirety of the text as context for each question. Instead, some sort of retrieval algorithm must be developed to select the specific chunk of text in the documents that contains the context that is most relevant to answer the user's query. That particular chunk of text is given to the LLM as additional input and the model then generates a response based on the question and the relevant context.  
    \\

    \noindent Below is a diagram of how a traditional RAG pipeline works. First, the user asks the model a question, for example for financial-based Q-A, they might ask "What is the FY2022 capital expenditure amount (in USD millions) for Company X?" This question is then converted into embeddinga or numerical representations of the text. In order to find the relevant context, we do a similarity search (such as cosine similarity) between the query embeddings and the embeddings contained in the knowledge hub. The knowledge hub is created from the set of documents we want to chat with. In this case, the documents might be a set of financial documents about various companies. After processing the unstructured documents in some way, the documents are divided into chunks and each chunk is converted to embeddings in the form of a vector. All these embeddings are stored in some sort of database, either locally or in the cloud and they consist of the knowledge hub. The algorithm retrieves the relevant chunk from this hub, and the chunk that is the most similar to the user query is chained to it and given to the LLM as context. 

    \begin{figure}
    \centering
    \includegraphics[width=0.5\linewidth]{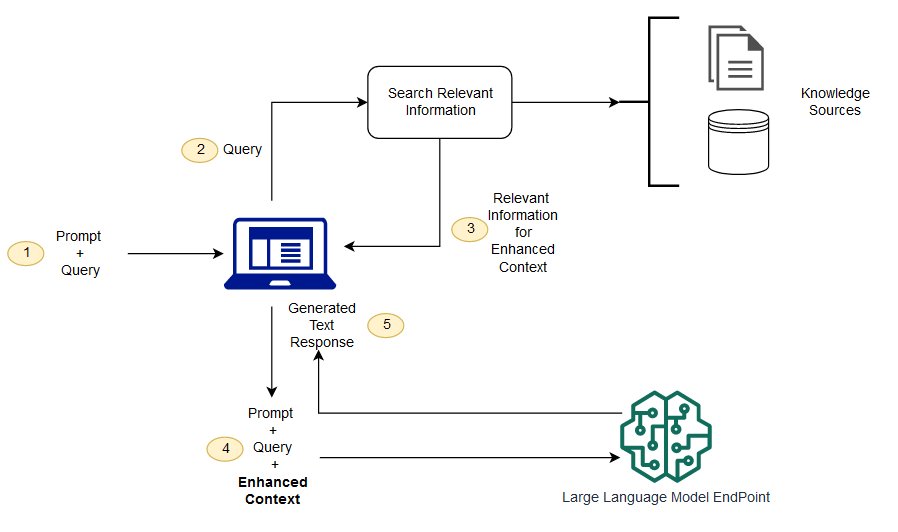}
    \caption{Retreival Augmented Generation Architecture}
    \end{figure}

    \subsection{Limitations of Current RAG Pipelines}   
    The traditional RAG architecture faces many limitations that impact its effectiveness in knowledge-intensive and domain-specific natural language processing tasks. Most RAG pipelines break the document into uniform chunks without any regard for the structure and content of the document. \footnote{Antonio Jimeno Yepes, Yao You, Jan Milczek, Sebastian Laverde, and Renyu Liu. Financial Report Chunking for Effective Retrieval Augmented Generation, 2024. arXiv:2402.05131.} Each chunk is made up of the same number of words or tokens and perhaps some overlap between chunks for context. The retriever will usually return the top k most similar chunks, and each chunk will be of roughly equal size. However, this approach does not account for the nature of the document at all and can lead to critical oversight and information loss in certain situations. For example, if the answer to the question stretched across a few different sections that was not necessarily the most semantically similar chunks, a traditional RAG pipeline would not return the most ideal chunk. 
    \\
    
    \noindent Semantic search via cosine similarity, a core component of RAG, also exhibits challenges such as retrieving irrelevant or opposing information, indicating the model’s sensitivity to language nuances and the potential for unexpected results. The most similar chunks are not necessarily the most relevant chunks, however many RAG pipelines make this assumption. Additionally, most times, standard embedding algorithms do not have any domain-specific knowledge and therefore, they overlook nuances of certain words or phrases in a specific field. \footnote{Sthanikam Santhosh. How to improve RAG (Retrieval Augmented Generation) performance, 2024. https://medium.com/@sthanikamsanthosh1994/how-to-improve-rag-retrieval-augmented-generation-performance-2a42303117f8}
    \\
    
    \noindent Basic chunking strategies and similarity search may have good results when working with a single long text as the knowledge base such as a book or article. However, in most cases, we are interested in the ability to chat with multiple different kinds of documents, which often have more complicated structures like headings and tables. When all the chunks are treated equally and thrown into a single vector database, critical information is lost. For example, the set of documents might be the same type of financial reports for various different companies or over the course of several years. Each chunk of text, depending on where in the document,  would not have contained information like what company it is about or what year with respect to certain metrics. This can lead to potentially incorrect and outdated information provided as the similarity search could return outdated or irrelevant chunks since it disregarded additional data about the text and documents.  
    \\
    
    \noindent Traditional retrieval algorithms lack the nuance and reasoning that is required for higher-level question answering on multiple different documents. Since it is usually just a similarity search based on the original question, the algorithm would not go through the same logical steps a human would in order to find the relevant section. For example, if you ask a financial analyst to calculate metric x, they would go through a multi-step reasoning process of knowing how to calculate the metrics and knowing where in the document that information is contained. Such complexities cannot be captured all the time with a similarity search with the question. The query itself might not have all the information needed to indicate to the model where the most relevant chunk is, but rather additional information and logical steps would be required.

    \section{Techniques to Improve Retrieval}
    Given the various limitations of current and standard RAG Pipelines, there is clearly a need to develop and implement different strategies in order to address these concerns. Throughout this section, we will explore a variety of different techniques that are aimed to optimize RAG performance for the use case of answering questions and chatting with various financial documents. 

    \subsection{Chunking Techniques}
    The chunking method is a critical aspect of the retrieval process at it determines the nature of the context that will be returned by the algorithm. Ineffective chunking strategies can lead to incomplete context in the chunks, or on the other extreme, too much irrelevant and unneeded information included in each chunk. When choosing a fixed chunk size, as most RAG pipelines do, it runs the risk of including too much or too little information for each query asked. The ideal approach will depend on the nature of the documents in the knowledge hub as well as the length of the queries the model expects and how much context is required to answer them. \footnote{Pinecone. Chunking Strategies, 2023. https://www.pinecone.io/learn/chunking-strategies/}
    \\ 
    
    \noindent Recursive Chunking is an example of a more adaptable chunking strategy that uses other indicators and rules like punctuation to make chunking more dynamic. While the chunks will still be of relatively equal size, the additional parameters will ensure that the chunks are not cut mid-sentence, for example. Through using Python libraries like Spacy and NLTK, we can use more sophisticated sentence-spitting techniques that use natural language processing techniques to be more aware of the context of the document. 
    \\
    
    \noindent When it comes to financial reports, these documents are generally quite long and contain more complicated structures like tables. Documents of the same nature, such as 10-ks also generally follow a specific format that can be indicated by the headings and subheadings of the documents. Due to the special nature of these financial reports, a version of element-based chunking that takes such facts into account could help to further improve retrieval. A paper by Unstructured proposed a novel way of breaking up documents into manageable sizes via element-based chunking.  \footnote{Antonio Jimeno Yepes, Yao You, Jan Milczek, Sebastian Laverde, and Renyu Liu. Financial Report Chunking for Effective Retrieval Augmented Generation, 2024. arXiv:2402.05131.} - if a title element is found, a new chunk is started and if a table element is found, a new chunk is started, preserving the entire table. Such strategies will help ensure that each context will return just enough context to effectively answer the user's question. 
    \\ 
    
    \noindent Other chunking methods include leveraging NLP techniques to implement recursive, character based chunking or semantic chunking. The nature of semantic or recursive chunking preserves the context of the document much better than character based or uniform chunking methods but become more complex to implement. A step about this is using something called agentic chunking where another LLM is prompted to break up the text in the best way possible. While this methods would seem to have the best contextual understanding, it is the most computationally expensive and often runs into the issue of exceeding the token size limit for the embedding models.

    \subsection{Query Expansion}
    Query Expansion, or query transformation refers to the idea of changing the question fed into the RAG pipeline to find relevant chunks based on more than the user's original question. The idea behind this is that oftentimes, the user's question may not explicitly contain all the information to indicate the algorithm where in the document it should look for the context, especially it is via a cosine similarity search. Sometimes, the user's question alone can mislead the algorithm to look in an incorrect area and therefore return an irrelevant chunk. For more complicated questions and documents, more reasoning steps are required to know where to look and basic RAG pipelines lack this completely. 
    \\
    
    \noindent A technique to somehow emulate the logic a human analyst might do with retrieval is to utilize Hypothetical Document Embeddings (HyDE). With HyDE, instead of just doing a similarity search with just the user's original query, it uses an LLM to generate a theoretical document when responding to a query and then does the similarity search with both the original question and hypothetical answer. \footnote{ AI Planet. Advanced RAG: Improving Retrieval using Hypothetical Document Embeddings (HYDE), 2023.  https://medium.aiplanet.com/advanced-rag-improving-retrieval-using-hypothetical-document-embeddings-hyde-1421a8ec075a}. This technique has been shown to outperform standard retrievers and eliminate the need for custom embedding algorithms, but can occasionally lead to incorrect results as it is dependent on another LLM for additional context.\footnote{Luyu Gao, Xueguang Ma, Jimmy Lin, Jamie Callan. Precise Zero-Shot Dense Retrieval without Relevance Labels, 2022. arXiv:2212.10496.}

    \begin{figure}[h!]
    \centering
    \caption{ illustration of HyDE \protect\footnotemark}
    \includegraphics[scale=0.45]{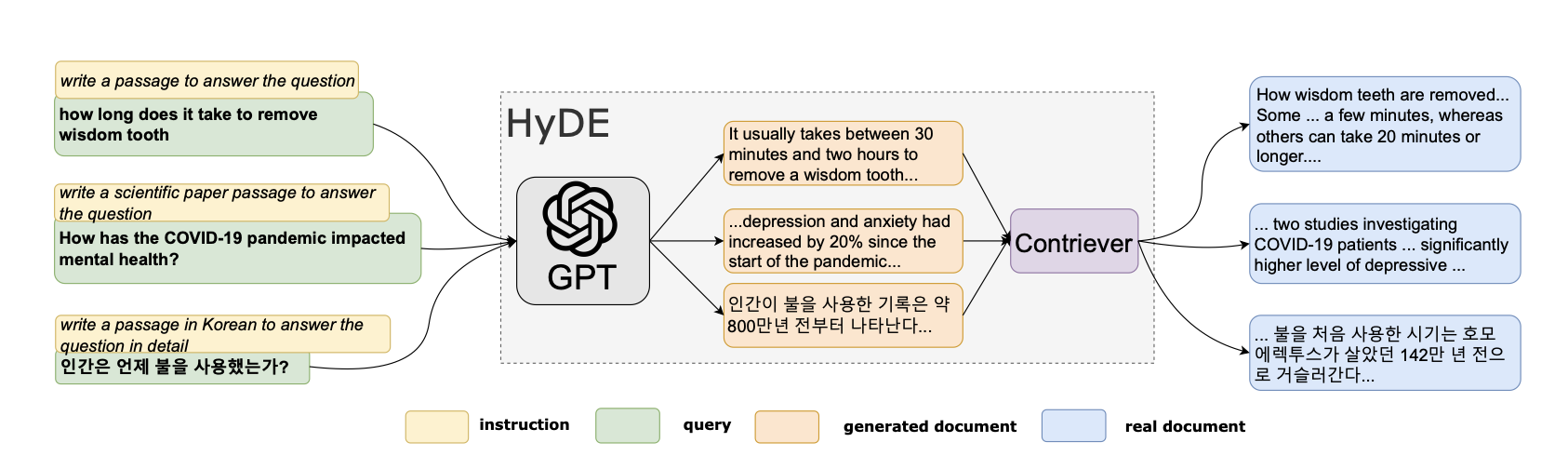}
    \end{figure}     \footnotetext{Luyu Gao, Xueguang Ma, Jimmy Lin, Jamie Callan. Precise Zero-Shot Dense Retrieval without Relevance Labels, 2022. arXiv:2212.10496.}

    \subsection{Metadata Annotations and Indexing}
    When interacting with several different documents, there might be key data points within the metadata that standard retrieval algorithms completely miss out on. A struggle we had faced previously when trying to chat with multiple documents, is that original RAG pipelines would confuse chunks of different documents, so to overcome this, separate vector databases were created for each document. However, this is not practical for large-scale applications and if the user wants to switch between documents or chat with multiple simultaneously. Metadata annotations is a method to overcome this hurdle and enhance retrieval further. 
    \\
    
    \noindent The nature of metadata annotations should be considered with the chunking strategies used. For example, since we will chunk the documents based on elements, details about each element can be included in the annotations. For example, if a table is created in a separate chunk, the metadata can include what kind of table it is (ex. income statement, cash flow statement, etc). Additionally, it is common to add summaries and representative keywords to metadata annotations for additional context.\footnote{Akash Mathur. Advanced RAG: Optimizing Retrieval with Additional Context Metadata using LlamaIndex, 2024. https://akash-mathur.medium.com/advanced-rag-optimizing-retrieval-with-additional-context-metadata-using-llamaindex-aeaa32d7aa2f} 

    \subsection{Re-ranking Algorithms}
    With standard RAG pipelines, one can specify the number of documents or chunks the algorithm should return and therefore how many chunks should be fed as context to the input query. Generally, the top 1 or 2 chunks are included as context, and these are the best results given a cosine similarity or k-nearest neighbors search. However, such algorithms will give the most similar chunks, which might not correspond to the most relevant chunks for context.  
    \\
    
    \noindent Re-ranking algorithms is a method to prioritize the relevance over the similarity of the chunks. Essentially, a method like cosine similarity might rank the top 10 chunks, but a separate algorithm will re-rank the algorithms to be based on relevance, and then the top one or two chunks after the re-ranking will be augmented as context to the input query. Cohere's re-ranking algorithm is a popular one and it along with others uses additional machine learning and natural language processing techniques to further evaluate relevance beyond a similarity search. In this paper we utilized a cross-encoder model to score how relevant each retreived chunk was to the question and re ranked the chunks based off the scores. The nature of the architecture suggests that cross encoder models have higher accuracy than bi-encoder models for this specific use case. \footnote{Cohere. Rerank, 2024. https://cohere.com/rerank}
    
    \subsection{Fine-tuning Embedding Algorithms}
    Embedding algorithms are what convert text into numerical representations, and play a crucial role in RAG pipelines. It is possible to fine-tune embedding algorithms based on domain-specific knowledge to enhance retrieval in that specific domain. Embeddings can also be dynamic where they adapt when words have slightly different meanings based on the context. OpenAI's embeddings are an example of dynamic embeddings that change based on the context. Domain-specific fine-tuning, however, requires access to datasets that consist of queries, the corpus of text, and relevant documents about that specific domain. In this paper, we are focusing on zero-shot methods to improve retrieval that require no additional training data. However future work will investigate how additional annotations/data can possible improve RAG via fine tuning embedding algorithms. \footnote{Yunfan Gao, Yun Xiong, Xinyu Gao, Kangxiang Jia, Jinliu Pan, Yuxi Bi, Yi Dai, Jiawei Sun, Qianyu Guo, Meng Wang, and Haofen Wang. Retrieval-Augmented Generation for Large Language Models: A Survey, 2024. arXiv:2312.10997.}

    \section{Evaluation}

    When it comes to evaluating RAG systems, the two main components are evaluating the model's ability to retrieve context as well as its ability to answer questions based on the context. In the datasets, we have access to ground truth context and answers developed by human financial analysts. Analysis of the model's responses with respect to these ground truth answers is referred to as structured evaluation, while evaluation without these ground truth answers is unstructured evaluation. Structured evaluation is more robust in indicating how accurate the model is, however, unstructured evaluation can be useful to tell us about the quality of the chunk and answer and is needed in scenarios when one does have access to ground truth data as is the case most of the time. 

    \subsection{Retrieval Quality}
    The primary way to test the retrieval quality with structured data is with page-level and paragraph-level accuracy. Since in the data, we have access to the entire document and the section a human analyst referred to, we compare that section to the chunks returned by the retrieval algorithm. If both the reference context and algorithm context are on the same page, it will have a high retrieval accuracy, a similar idea for paragraph-level accuracy. 
    \\
    
    \noindent For unstructured data, we can evaluate the retrieved chunk using Context Relevance, as defined by the RAGAS Framework. \footnote{ Shahul Es, Jithin James, Luis Espinosa-Anke, Steven Schockaert. RAGAS: Automated Evaluation of Retrieval Augmented Generation. 2023, arXiv:2309.15217}For this metric, an LLM is asked to count the number of sentences from the context that are relevant to answering the question. The ratio of the number of extracted sentences to the total number of sentences in the context is defined as the context relevance score. This score penalizes redundant information and rewards chunks that have a majority of sentences that provide useful information to answer the question. 

    \subsection{Answer Accuracy}
    For structured evaluation of question-answering, we evaluate the accuracy of the model's answer to the ground truth answers from the dataset. We use standard metrics including BLEU and Rouge-L Score, as well as cosine similarity. However, these scores are not ideal for comparing the semantic meaning of two different answers, and can therefore be misleading. Therefore, we also use a version of LLM evaluation where we use a specific prompt for a model like GPT and ask it to evaluate the accuracy of the candidate's response given a ground truth answer. 
    \\
    
    \noindent We can further evaluate the quality of the generator's answer with respect to the retrieved context in an unstructured manner. These metrics measure the quality of the model's response given the context. Answer faithfulness is another metric presented by the RAGAS framework that can be used which measures how well the model's answer is grounded in the given context. It is measured by dividing the number of statements in the answer that are supported by the context by the total number of sentences in the answer. This score can also be seen as a way of evaluating whether a not a model is hallucinating information without the need for ground truth answers. A high faithfulness score means that almost all the sentences in the generated answer are supported by the context, meaning that the model did not make up any additional information outside of the documents. 

    \section{Results}
    To assess the performance of a model with RAG, we utilized the FinanceBench benchmark for question and answering models developed by Patronus AI. This benchmark consists of 10,231 questions about publicly traded companies in the USA, covering a range of financial documents such as 10Ks, 10Qs, 8Ks, and Earnings Reports released between 2015 and 2023. However only 150 rows were publicly avaiable, so that is what was used in this study 
    \\
    
    \noindent Each entry in the FinanceBench dataset contains a question (e.g., “What Was AMCOR’s Adjusted Non-GAAP EBITDA for FY 2023”), answer (e.g., “AMCOR’s Adj. EBITDA was 2,018 million USD in FY 2023”), evidence string (containing information needed to verify the answer), and a page number from the relevant document. This dataset provides a comprehensive evaluation of the RAG model's ability to retrieve and generate accurate answers to financial questions, highlighting LLM's potential for real-world applications in financial analysis and decision-making. 
\\

   \noindent The FinanceBench dataset was tested for the general RAG model called base RAG. In the base case, the RAG model was not given additional context, meaning the model had to retrieve context information on its own. On the contrary, another experiment was tested, we will call this Fake RAG. Fake RAG uses the same LLM model but is given the correct context to answer the original question, specifically using context information from FinanceBench. We also have results for the whole dataset for the query expansion techniques and a reranking approach using a cross encoder. There are some initial results for testing different chunking techniques, however because of compute limitations as well as it exceeding the token limit of the model, we could not get a comprehensive metric of the whole dataset and therefore cannot compare it directly with the other methods. The preliminary results for those chunking methods will be available in the github repository for anyone interesting in seeing the initial results and code. While we had started to experiment with some metadata tagging techniques, there was not a clear way to integrate such with the existing pipeline so results for that technique are not currently available, but the corresponding repository will be updated when we have run successful experiments.  
\\ 
\\
   As you can see from the plot and table, the "fake rag" approach performed the best with an average LLM evaluation score of 0.573. This was expected to have the best results as the LLM was given the context that was required to answer the question correctly. However, as you can see from the table, this method had the highest standard deviation or variation in the score. When looking through the data, it seems that when given the right context, the LLM answered the question correctly, or got it completely wrong. So while giving the correct context significantly improves the accuracy of the model, there are still limitation swith the LLM itself, even when using the state of the art generator like GPT-4o, when answering highly complex and domain specific questions. 
\\
\\
   The method with the lowest accuracy was the base rag case, with an average LLM evaluation score of 0.204. The same LLM and prompt template was used for each method, only the way the context was retrieved differed. This means that the difference in the context fed into the model contributed to a gap in accuracy of over 35 percent, underscoring the importance of retrieving correct texts, especially for longer more complex documents. The zero shot methods of query expansion and re-ranking seem to have some improvement on the accuracy with LLM evaluation scores of 0.24 and 0.256 respectively. While this is no where near the accuracy of the fake rag results when it is fed the correct context, its improvement over the base rag case is not insignificant.
   From this, we can conclude that zero shot methods to enhance retrieval might be in the step in the right direction to improve accuracy, but not nearly enough to make robust systems with very high accuracy. In order to to achieve that goal, other techniques, perhaps those that require some additional training or fine tuning of some sorts can be utilized. 
\\
\\
\begin{figure}
    \centering
    \includegraphics[width=0.5\linewidth]{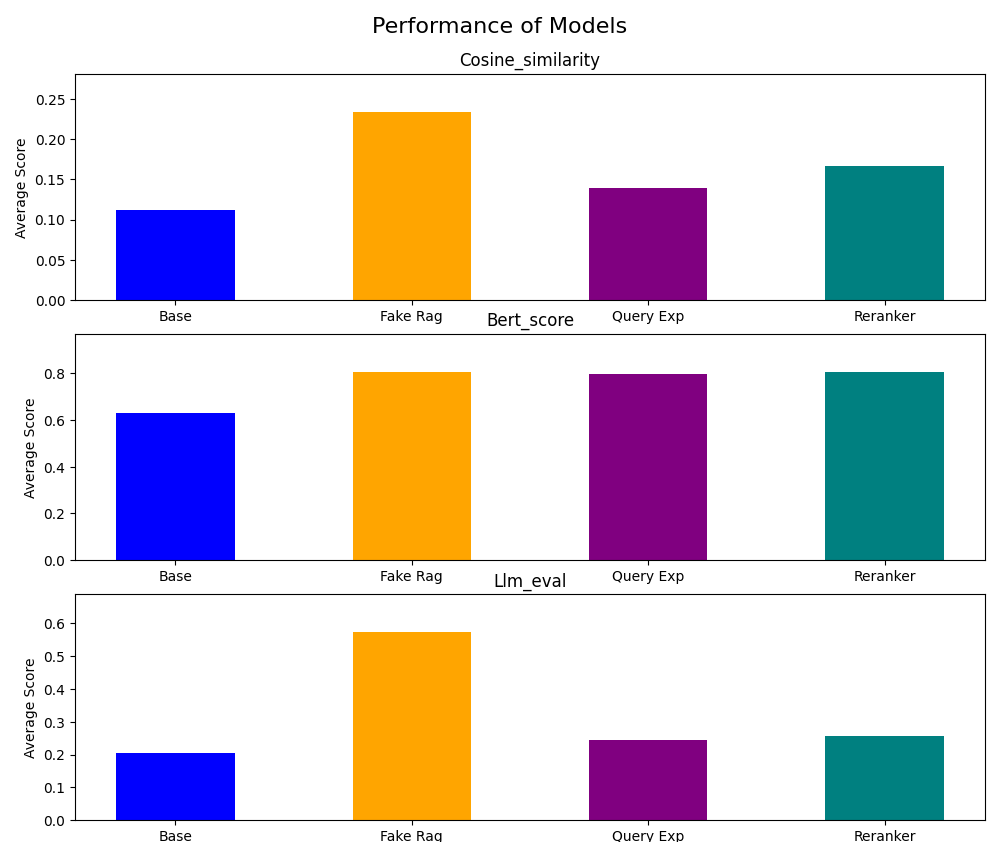}
    \caption{Aggregate Results Plot}

    \centering
    \includegraphics[width=0.5\linewidth]{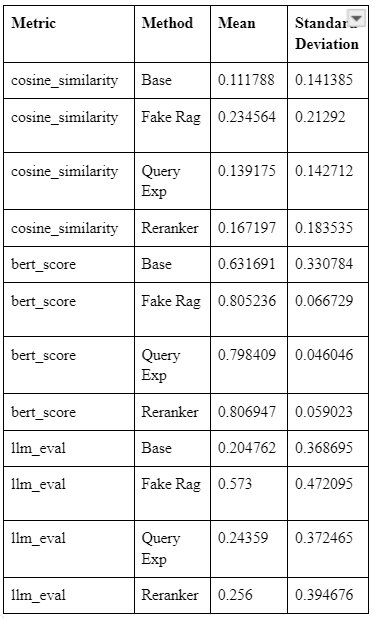}
    \caption{Aggregate Results Table}
\end{figure}

    \section{Impact}
    \noindent In this experiment, we aim to address the key limitations of current RAG pipelines, some of these concerns have specific techniques to address them, while other are more difficult to resolve. As research in this field continues, more ideas will emerge to make these system more accurate and robust. They key concerns we have identified can be viewed as three difference categories. The first concern is that relevant context is often located in multiple different locations, and current algorithms only retrieve 1 or 2 chunks. Even if these algorithms retrieve the top 5 - 10 chunks, these are not necessarily the current location of the most relevant sections that a human analyst, would know to look. The second key concern is that current RAG pipelines assume that similarity equivalence to relevance while this is not correct. The final concern is that current RAG pipelines chunk things in uniform sections that disregard document structures can can leave certain parts incomplete.

    \section{Conclusion and Next Steps}
    \noindent By improving the performance of retrieval we improve the quality of the entire system for document based question-answering tasks. Through retrieving the right chunk, we not only provide better citations, but from there we can also provide better answers to questions. Such underscores the importance of robust retrieval algorithms, as without the right context, even the best generators will give incorrect answers. We focused on use cases for RAG systems in the financial domain, primarily question-answering on 10-K documents. However, the same concept are applicable across industries and applications in healthcare, and legal tech, as it provides a systemic framework to improve RAG pipelines for domain specific tasks. We also employ a variety of evaluation metrics to show that these enhanced rag techniques overcame some of the limitations. While each metric alone cannot provide a complete overview, all together they give a sense of performance of the system, with and without the availability of structured data. 
    \\

    \noindent While the techniques explored in this paper address certain key limitations in current RAG pipelines, there are for other areas in which more could be done to improve performance. In particular, a key struggle for retrieval algorithms is when the context is across multiple parts in the document, and the model must find and use all these sections in order to formulate an answer. Implementing knowledge graphs could help retrieval systems do this by giving them a specific set of instructions of how to retrieve questions and answer questions depending on the question asked. Another technique to explore is fine tuning embedding algorithms based on data labeled by the user so that the model can better understand the nuance of each word and phrase in a domain specific context. 

 \pagebreak 
\section{References}


\small
\begin{enumerate}

    \item Shahul Es, Jithin James, Luis Espinosa-Anke, Steven Schockaert. RAGAS: Automated Evaluation of Retrieval Augmented Generation. 2023, arXiv:2309.15217.

   \item Oded Ovadia, Menachem Brief, Moshik Mishaeli, and Oren Elisha.
    Fine-Tuning or Retrieval? Comparing Knowledge Injection in LLMs, 2023. arXiv:2312.05934.

   \item Antonio Jimeno Yepes, Yao You, Jan Milczek, Sebastian Laverde, and Renyu Liu. Financial Report Chunking for Effective Retrieval Augmented Generation, 2024. arXiv:2402.05131.
    
    \item Sthanikam Santhosh. How to improve RAG (Retrieval Augmented Generation) performance, 2024. https://medium.com/@sthanikamsanthosh1994/how-to-improve-rag-retrieval-augmented-generation-performance-2a42303117f8

    \noindent \item Pinecone. Chunking Strategies, 2023.
    https://www.pinecone.io/learn/chunking-strategies
    
     \item AI Planet. Advanced RAG: Improving Retrieval using Hypothetical Document Embeddings (HYDE), 2023.  
     https://medium.aiplanet.com/advanced-rag-improving-retrieval-using-hypothetical-document-embeddings-hyde-1421a8ec075a

     \item Luyu Gao, Xueguang Ma, Jimmy Lin, Jamie Callan. Precise Zero-Shot Dense Retrieval without Relevance Labels, 2022. arXiv:2212.10496.

    \item Akash Mathur. Advanced RAG: Optimizing Retrieval with Additional Context Metadata using LlamaIndex, 2024.
    https://akash-mathur.medium.com/advanced-rag-optimizing-retrieval-with-additional-context-metadata-using-llamaindex-aeaa32d7aa2f
    
     \item Cohere. Rerank, 2024. https://cohere.com/rerank

    \item  Yunfan Gao, Yun Xiong, Xinyu Gao, Kangxiang Jia, Jinliu Pan, Yuxi Bi, Yi Dai, Jiawei Sun, Qianyu Guo, Meng Wang, and Haofen Wang. Retrieval-Augmented Generation for Large Language Models: A Survey, 2024. arXiv:2312.10997.

    \item Spurthi Setty. Rerankertestresultstest.csv, 2024. https://github.com/n78/Anote

    \item Henry Toll. Chunkingevaluationresults.csv, 2024. https://github.com/n78/Anote

    \item Spurthi Setty, Henry Toll. Queryexpansionresults.csv, 2024. https://github.com/nv78/Anote

\end{enumerate}

\end{document}